\documentclass[10pt,aps,prl,superscriptaddress,twocolumn,preprintnumbers,nofootinbib]{revtex4-2}
\usepackage{amsmath,amssymb,mathtools}
\usepackage{booktabs}
\usepackage{graphicx}
\usepackage{lineno}
\usepackage{hyperref}
\usepackage{color}
\hypersetup{
	colorlinks=true,
	linkcolor=blue,
	citecolor=blue,
	urlcolor=blue,
}

\newcommand{\sect}[1]{%
	\bigskip\noindent%
	{\bfseries\upshape\rmfamily\boldmath{#1}}---%
	\ignorespaces%
}

\newcommand{\ycut}{y_{\mathrm{cut}}}
\newcommand{\alphas}{\alpha_{\mathrm{s}}}

\begin{document}
	
	
	\preprint{IPPP/25/10, ZU-TH 09/25, MCNET-25-03}
	
	\title{Jet rates in Higgs boson decay at third order in QCD}
	
	\author{Elliot Fox}
	\affiliation{%
		Institute for Particle Physics Phenomenology, Department of Physics, University of Durham, Durham, DH1 3LE, UK}%
	\author{Aude Gehrmann-De Ridder}
	\affiliation{%
		Institute for Theoretical Physics, ETH, 8093 Z\"{u}rich, Switzerland}%
	\affiliation{%
		Physik-Institut, Universit\"{a}t Z\"{u}rich, 8057 Z\"{u}rich, Switzerland}%

	\author{Thomas Gehrmann}
	\affiliation{%
		Physik-Institut, Universit\"{a}t Z\"{u}rich, 8057 Z\"{u}rich, Switzerland}%
	\author{Nigel Glover}
	\affiliation{%
		Institute for Particle Physics Phenomenology, Department of Physics, University of Durham, Durham, DH1 3LE, UK}%
	\author{Matteo Marcoli}
	\affiliation{%
		Institute for Particle Physics Phenomenology, Department of Physics, University of Durham, Durham, DH1 3LE, UK}%
	\author{Christian T.~Preuss}
	\affiliation{%
		Department of Physics, University of Wuppertal, 42119 Wuppertal, Germany}%
	
	\begin{abstract}
		We compute the production rates for two, three, four and five jets in the hadronic decay 
		of a Higgs boson in its two dominant decay modes to 
		bottom quarks and gluons  to third order in the QCD coupling constant.  
		The five-, four- and three-jet rates are obtained from a next-to-next-to-leading order (NNLO) calculation of Higgs decay to three jets, while the two-jet rate is inferred at next-to-next-to-next-to-leading order (N$^3$LO) from the inclusive decay rate. Our results show distinct differences in the dependence of the jet rates on the jet resolution parameter between the two decay modes, supporting the aim of discriminating different Higgs boson decay channels via classic QCD observables. 	\end{abstract}
	
	\maketitle
	
	
	
	Since the discovery of the Higgs boson in 2012~\cite{ATLAS:2012yve,CMS:2012qbp}, its properties have been extensively investigated at the LHC. The understanding of the Higgs boson interactions with other fundamental particles is crucial to shed light onto the electroweak symmetry breaking, the flavour structure of the Standard Model and potential new physics phenomena.
	
	The Higgs boson couplings to massive gauge bosons, top quarks, tau leptons and muons have already been extracted from measurements of Higgs production and non-hadronic decay cross sections~\cite{ATLAS:2022vkf,CMS:2022dwd}. On the other hand, the investigation of the hadronic decay modes which probe the coupling of the Higgs boson to light quarks via the Yukawa interaction and to gluons via a heavy quark loop is limited at hadron colliders by the large QCD background. To date, only the dominant decay mode to bottom quarks has been observed~\cite{ATLAS:2018kot,CMS:2018nsn}. 
	
	Future lepton colliders such as the FCC-ee~\cite{FCC:2018byv,FCC:2018evy} or the CEPC~\cite{CEPCStudyGroup:2018ghi} are intended to operate as so-called `Higgs factories' for the collection of a very large number of events with a Higgs boson in the final state, via the Higgsstrahlung process $e^+e^-\to ZH$ and vector-boson-fusion $e^+e^-\to \ell\bar{\ell}H$. Due to their significantly cleaner environment, free from QCD initial-state radiation and multi-parton interactions, such colliders offer the possibility of reaching unprecedented resolution on the hadronic decays of the Higgs boson. 
	
While the inclusive branching fractions of Higgs bosons to bottom quarks and to gluons are known to 
fourth order in perturbative QCD~\cite{Baikov:2005rw,Baikov:2006ch,Davies:2017xsp,Herzog:2017dtz}, 
predictions for  more differential observables such as jet rates or event-shape distributions are restricted to 
lower orders. Previous precision 
calculations  
focused on  the decay to bottom quarks, where 
three-jet production~\cite{Mondini:2019vub} is known to next-to-next-to-leading order (NNLO) in QCD  
and two-jet production~\cite{Mondini:2019gid} to third order (N$^3$LO), with quark mass corrections known up to NNLO~\cite{Bernreuther:2018ynm}. Comprehensive studies of 
hadronic event-shape distributions 
in both types of decay modes have been performed at next-to-leading order (NLO) in QCD 
for shape variables related to three-jet~\cite{Coloretti:2022jcl,Gehrmann-DeRidder:2024avt} 
and four-jet~\cite{Gehrmann-DeRidder:2023uld} final states, indicating their potential for discriminating 
different Higgs boson decay modes~\cite{Gao:2016jcm,Coloretti:2022jcl,Knobbe:2023njd} in combination with 
final-state flavour identification~\cite{CampilloAveleira:2024fll}. 
	
In this Letter, we compute for the first time the QCD predictions for the decay 
of a Higgs boson to three jets to NNLO 
and to two jets to N$^3$LO for all types of hadronic decay modes.  
We study the phenomenological impact of the corrections and expose the differences between the decay modes, 
thereby preparing for future precision studies of hadronic Higgs boson decays.

\begin{figure}
      \includegraphics[width = 0.45\textwidth]{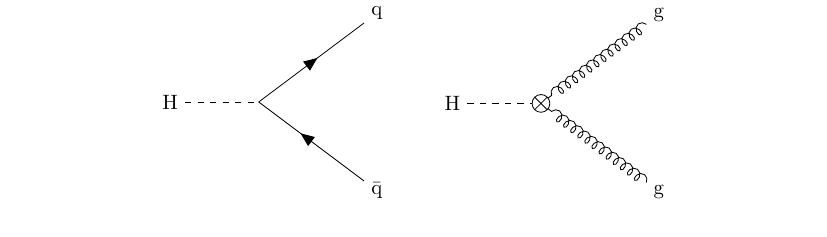}
		\caption{Feynman diagrams of the two decay categories of a Higgs boson decaying into two jets via the Yukawa interaction (left) and the effective gluon vertex (right).}
		\label{fig:h2j_diags}
	\end{figure}

	
	We consider the two dominant decay modes of Higgs bosons to 
	hadrons, displayed in Figure~\ref{fig:h2j_diags}.
	In the first case, the Higgs boson couples to a quark-antiquark pair via Yukawa interaction. Although we consider non-vanishing bottom-quark Yukawa couplings, light quarks, including the $b$ quark, are treated as massless in the whole calculation.  In the second case, the Higgs boson decays into a pair of gluons via a heavy quark loop, which produces an effective vertex in the limit of an infinite heavy-quark mass~\cite{Wilczek:1977zn,Shifman:1978zn,Inami:1982xt}. 
The Born-level two-parton decay process corresponds to $\alphas^0$ for the Yukawa interaction mode 
and to $\alphas^2$ for the 
effective Higgs-to-gluons interaction mode. Perturbative higher-order corrections are defined relative to the coupling 
order of the Born process throughout.  
Due to their different chirality structure, the two decay modes do not interfere at any order in perturbation theory in 
the limit of vanishing bottom quark mass. We explicitly verified that the Born-level interference contributions for massive bottom quarks remain well below one percent for all observables considered here.

The N$^k$LO Higgs boson decay rate in the two decay modes reads: 
\begin{eqnarray}
		\label{eq:gamma_bb}\Gamma^{(k)}_{H\to b\bar{b}}&=&\Gamma^{(0)}_{H\to b\bar{b}}\left(1+\sum_{n=1}^k\alphas^n(\mu_R)C_{b\bar{b}}^{(n)}\right),\\
		\label{eq:gamma_gg}\Gamma^{(k)}_{H\to gg}&=&\Gamma^{(0)}_{H\to gg}\left(1+\sum_{n=1}^k\alphas^n(\mu_R)C_{gg}^{(n)}\right),
	\end{eqnarray} 
	where $\mu_R$ is the renormalisation scale,
	\begin{eqnarray}
		\label{eq:gamma0_bb}\Gamma^{(0)}_{H\to b\bar{b}}&=&\dfrac{y_b^2(\mu_R) m_H N_c}{8\pi},\\
		\label{eq:gamma0_gg}\Gamma^{(0)}_{H\to gg}&=&\dfrac{\lambda_0^2(\mu_R) m_H^3(N_c^2-1)}{64\pi},
	\end{eqnarray} 
	and
	\begin{equation}\label{eq:lambda0}
		\lambda_0^2(\mu_R)=\dfrac{\alpha_s^2(\mu_R)\sqrt{2}G_F}{9\pi^2}.
	\end{equation}
	The coefficients $C^{(n)}_{b\bar{b}}$ and $C^{(n)}_{gg}$ up to N$^4$LO are given for example in~\cite{Herzog:2017dtz}.
 		
Particles are clustered into jets using a jet algorithm with some pre-defined resolution parameter. We employ 
	the Durham jet algorithm~\cite{Catani:1991hj,Brown:1990nm,Brown:1991hx,Stirling:1991ds,Bethke:1991wk}, 
	which has been used most widely in jet studies in $e^+e^-$ annihilation. The resulting Higgs decay rates 
	for different multiplicity then depend on the value of the resolution parameter $\ycut$, and jet multiplicities can be 
	defined by normalising the Higgs decay rates to the total hadronic Higgs-boson decay rate. 
	 For both decay categories, we consider a Born-level configuration where the Higgs-boson decays to three partons and we compute the relative NNLO correction. 
	 
	The $n$-jet decay rate at relative ${\cal O}(\alphas^k)$ depends on the value of $\ycut$ and is denoted by $\Gamma_{X}^{(k)}(n,\ycut)$. Along with the three-jet rate $\Gamma^{(3)}_X(3,\ycut)$, by-products of the NNLO three-jet calculations are the four-jet rate at NLO, $\Gamma^{(3)}_X(4,\ycut)$ and the five-jet rate at LO, $\Gamma^{(3)}_X(5,\ycut)$.
	The normalised jet rate (or jet fraction) is defined with respect to the inclusive width at the same order,
	\begin{equation}
	R^{(k)}_{X}(n,\ycut) = \frac{\Gamma_{H\to X}^{(k)}(n,\ycut)}{\Gamma_{H\to X}^{(k)}},\,\,\text{with}\,X=gg,\,b\bar{b},\,
	\end{equation}
	such that the rates for all possible jet multiplicities at that order sum to unity for any value of $\ycut$,
	\begin{equation}
	\label{eq:unitarity}
		\sum_{n=2}^{k+2} R^{(k)}_{X}(n,\ycut) = 1.
	\end{equation}
	Since the Higgs boson total decay rate to hadrons is  known up to N$^4$LO~\cite{Baikov:2005rw,Baikov:2006ch,Davies:2017xsp,Herzog:2017dtz} in either mode, we can infer the two-jet fraction at N$^3$LO as a function of $\ycut$, $R^{(3)}_{X}(2,\ycut)$, from the knowledge of the three-, four- and five-jet fractions, as done in~\cite{Gehrmann-DeRidder:2008qsl} in the context of electron-positron annihilation and associated to $Z$ decays.

	The calculation involves matrix elements for the decay of a Higgs boson to up to five partons at 
	tree-level~\cite{DelDuca:2004wt,Anastasiou:2011qx,DelDuca:2015zqa,Mondini:2019vub,Mondini:2019gid}, up to four partons at one loop~\cite{Dixon:2009uk,Badger:2009hw,Badger:2009vh,Anastasiou:2011qx,DelDuca:2015zqa,Mondini:2019vub,Mondini:2019gid} and three partons at two loops. The two-loop Higgs-plus-three-parton amplitudes for an effective Higgs-to-gluons coupling were derived in~\cite{Gehrmann:2011aa} and implemented by a subset of the authors for the calculation of NNLO corrections to the production of a Higgs boson and a jet at hadron colliders~\cite{Chen:2014gva,Chen:2016zka}. The two-loop amplitudes for $H\to b\bar{b} g$ were originally derived in~\cite{Ahmed:2014pka,Mondini:2019vub}. We recomputed them, obtaining full numerical agreement with~\cite{Mondini:2019vub}.
	
	The computation of higher-order QCD corrections demands suitable techniques for the removal of infrared (IR) divergences. After renormalisation, loop amplitudes contain explicit IR singularities, while the emission of on-shell massless particles leads to the divergent behaviour of real-emission matrix elements in soft and collinear limits. These divergences only cancel when virtual and real corrections are  combined in physical predictions.	
	
	\begin{figure*}[t]
		\centering
		\label{fig:3jetNNLOa}\includegraphics[width=0.32\textwidth]{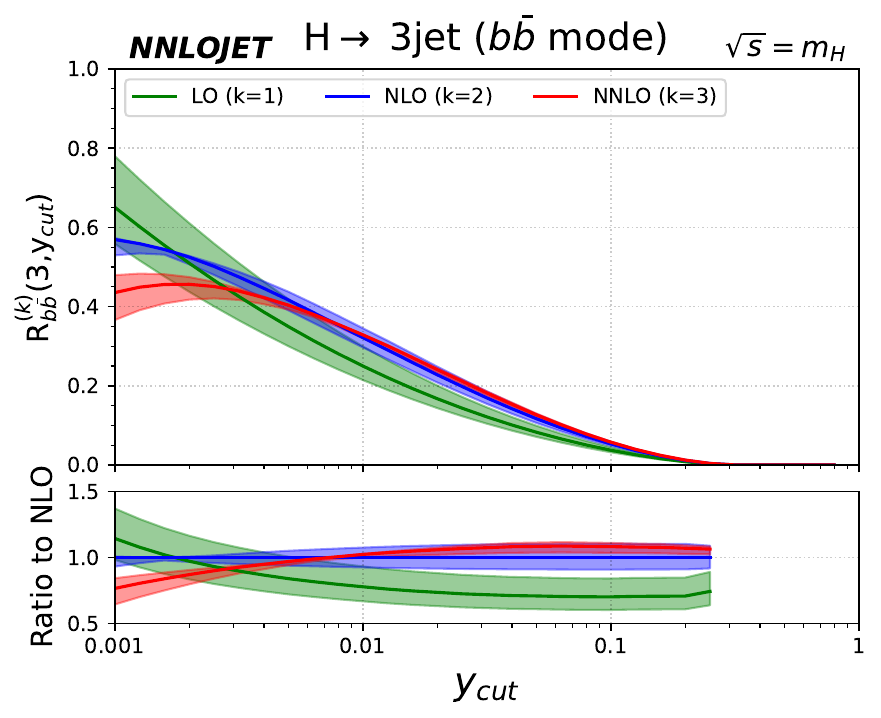}
		\label{fig:3jetNNLOb}\includegraphics[width=0.32\textwidth]{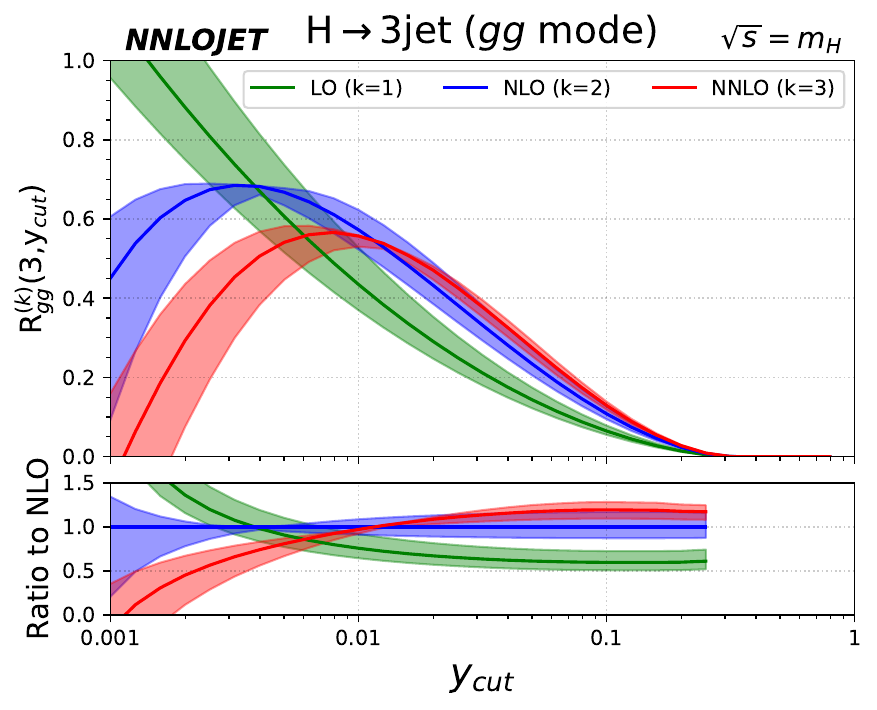}
		\label{fig:3jetNNLOc}\includegraphics[width=0.32\textwidth]{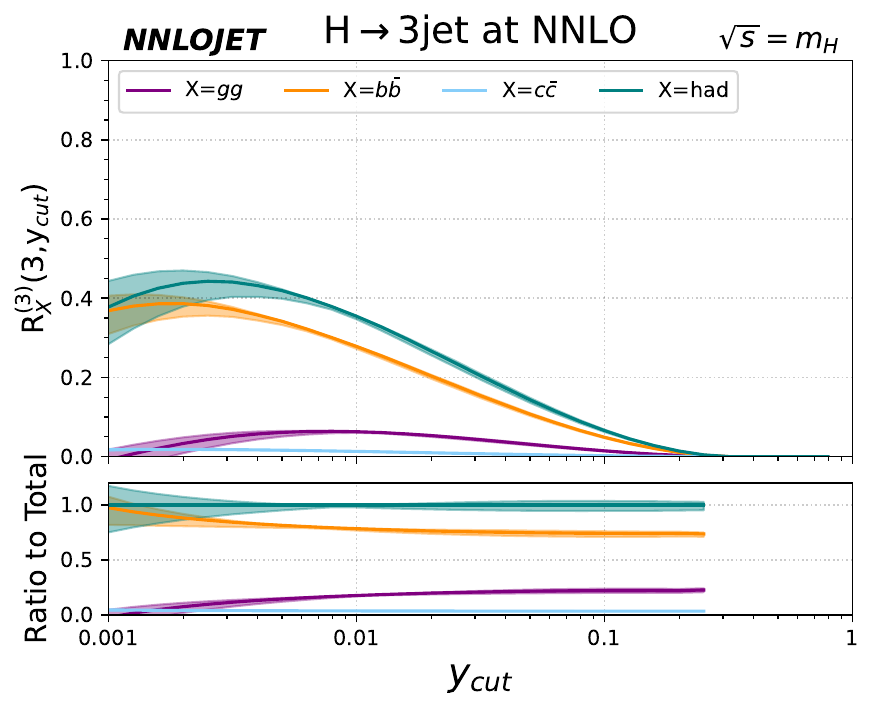}
		\caption{Normalised three-jet decay rate at LO (green), NLO (blue) and NNLO (red) for Higgs decay to bottom quarks (left), Higgs to gluons (centre), and the weighted sum of the different decay modes (right). The corresponding inclusive decay rate is used as normalisation (see text).}
		\label{fig:3jetNNLO}
	\end{figure*} 
		
	We rely on the antenna subtraction method for NNLO calculations in QCD~\cite{Gehrmann-DeRidder:2005btv,Currie:2013vh}, which employs \textit{antenna functions} to assemble counterterms for the removal of the divergent behaviour of real-emission matrix elements in infrared configurations. These counterterms are then analytically integrated over the inclusive phase space of the radiation and combined with the virtual corrections, subtracting their explicit infrared singularities. In particular, we implement the \textit{generalised antenna functions} for final-state radiation recently presented in~\cite{Fox:2024bfp}, constructed by the application of the algorithm described in~\cite{Braun-White:2023sgd,Braun-White:2023zwd} to build antenna functions from a target set of infrared limits. This newly proposed version of the antenna subtraction method offers significant advantages in terms of simplicity of the subtraction terms and computational efficiency. In particular, the improvements given by the generalised antenna functions scale with the number of external gluons, meaning that we expect more than an order of magnitude of computational speed-up in the gluonic decay mode, with respect to the traditional antenna subtraction method.

	 In~\cite{Fox:2024bfp}, the detailed validation of all the structures needed for the computation of the NNLO correction to three-jet production in electron-positron annihilation is discussed, and an identical setup is employed here for the Yukawa-mediated mode. The NNLO calculation in the gluonic mode presented in this Letter completes this implementation, now employing the full 
	set of final-state generalised antenna functions~\cite{Fox:2024bfp}.
	The IR-finite remainders are numerically integrated with the Monte Carlo event generator \textsc{NNLOjet}, suitably adapted for the implementation of generalised antenna functions. 
	
	We validated our LO and NLO results for the production of three and four jets against \textsc{Eerad3}~\cite{Gehrmann-DeRidder:2014hxk,Coloretti:2022jcl,Gehrmann-DeRidder:2023uld} and find very good agreement. For the Yukawa-induced mode we reproduce the results of~\cite{Mondini:2019vub,Mondini:2019gid}.
	
	We consider the decay of an on-shell Higgs boson with a mass of  $m_H=125.09 \, \text{GeV}$. We work in the $G_{\mu}$-scheme with constant electroweak parameters:
	\begin{eqnarray}
		G_F &=& 1.1664\cdot 10^{-5}\, \text{GeV}^{-2}\,, \nonumber\\
		m_Z &=&  \text{91.200}\, \text{GeV}\,,\nonumber
	\end{eqnarray}
	yielding a Higgs vacuum expectation value of \mbox{$v=\text{246.22}~\text{GeV}$}. The renormalisation scale is chosen to be $\mu_\mathrm{R}=m_H$, and we assess theory uncertainties by varying $\mu_R$ in the range $[m_H/2,2m_H]$.  	The QCD coupling is set to \mbox{$\alphas(m_Z)=0.11800$} and its scale evolution is performed with LHAPDF6~\cite{Buckley:2014ana}, yielding \mbox{$\alphas(m_H)=0.11263$}. 
	 
We specifically focus on the decay to bottom quarks and gluons to expose their differences, but also consider the combination of the two modes. To obtain phenomenologically viable 
results for the combination of the different decay modes, we also include the Yukawa decay to charm quarks, which is identical to the bottom Yukawa decay up to its overall normalisation. Moreover, we rescale $\lambda_0^2(\mu_R)$ in~\eqref{eq:lambda0} to account 
for the exact top, bottom and charm mass dependence of the one-loop Higgs-gluon-gluon vertex~\cite{Spira:1997dg}, including interference contributions. The combined result is normalized with respect to the total decay rate to hadrons at N$^k$LO:
	\begin{equation}
		\label{eq:gamma_tot}\Gamma^{(k)}_{H\to \text{had}}=\Gamma^{(k)}_{H\to b\bar{b}}+
	\Gamma^{(k)}_{H\to c\bar{c}}+	\Gamma^{(k)}_{H\to gg}.
	\end{equation}

	We take running 
	 $\overline{\mathrm{MS}}$ bottom and charm quark masses and Yukawa couplings
	 \mbox{$y_b(m_H)=m_b(m_H)/v=0.011309$}, \mbox{$y_c(m_H)=m_c(m_H)/v=0.0024629$}  and 
use the $\overline{\mathrm{MS}}$
	top mass \mbox{$m_t(m_H) = 166.48\,\text{GeV}$}.  The Yukawa couplings and the top-quark mass are evolved to different scales according to~\cite{Vermaseren_1997}.
	
	In the first two frames of Figure~\ref{fig:3jetNNLO}, we present the NNLO correction to the normalised three-jet rate for the two different decay modes. 
	To estimate theory uncertainties, the renormalisation scale is varied in a correlated manner between numerator and denominator.
	
	We observe that in both channels the NNLO correction is small for large values of $\ycut$, but is significant and negative in regions of lower $\ycut$. The three-jet fraction attains a maximum at a certain value of 
$\ycut$. Upon inclusion of the NNLO corrections, the height of this three-jet peak decreases and its location 
is shifted to the right. For the NNLO predictions, the three-jet peak lies at $\ycut\approx 0.002$ and at $\ycut\approx 0.007$ for the $b$-quark and the gluonic decay channel respectively, showing a first significant difference between the two modes.
	
	For values of $\ycut$ to the right of the peak, the NNLO result lies within the NLO band, indicating good perturbative convergence, and the relative size of the theory uncertainties is halved. On the other hand, to the left of the peak, the NLO and NNLO uncertainty bands do not overlap and the two curves exhibit sizeable differences. Indeed, for low $\ycut$ values one expects the emergence of large logarithmic contributions at each order in $\alphas$, which spoil the convergence of the perturbative series. In this region, all-order logarithmic resummation is required to obtain sensible predictions for the jet rates. One can consider the location of the three-jet peak as a reasonable estimate for the 
	onset of resummation effects 
	 in the respective channels. We observe that for the gluonic decay mode, this happens at values of $\ycut$ 
	 about three times as large as the ones for the Yukawa-induced channel.	\begin{figure*}[t]
		\centering
		\includegraphics[width=0.32\textwidth]{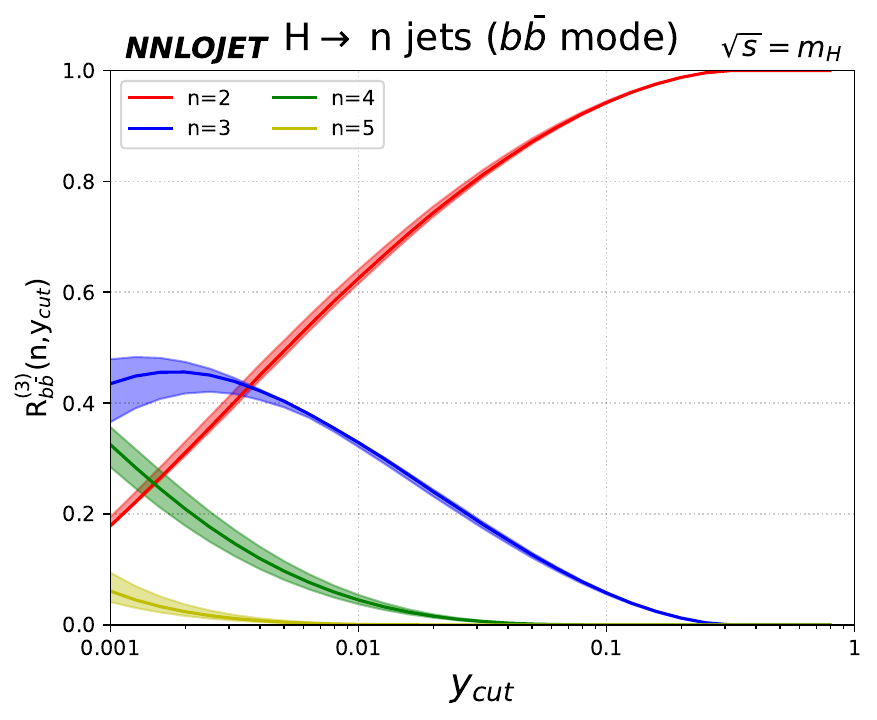}
		\includegraphics[width=0.32\textwidth]{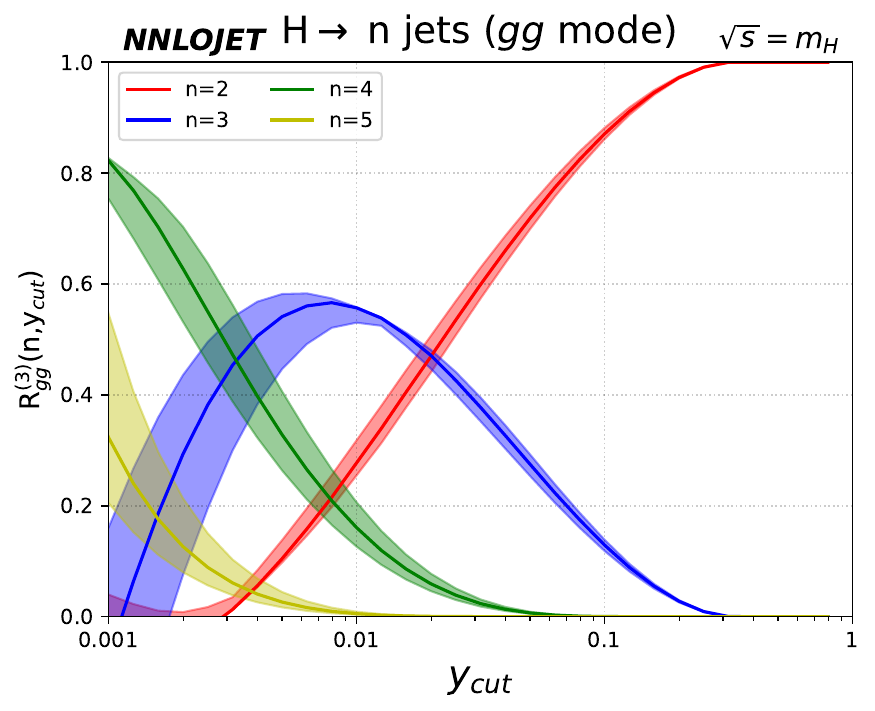}
		\includegraphics[width=0.32\textwidth]{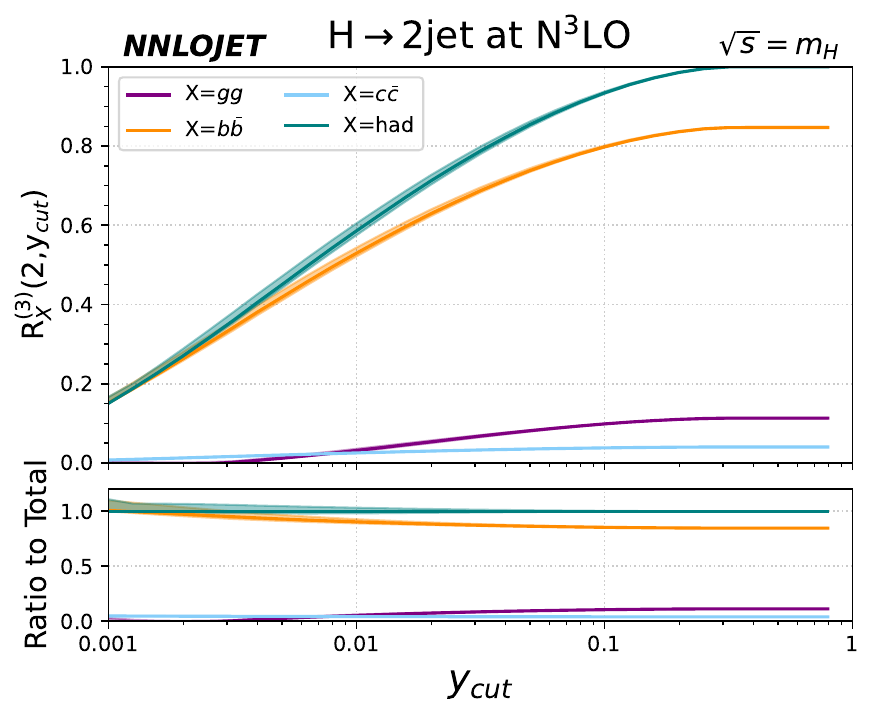}
		\caption{Fractional jet rates for the production of two (red), three (blue), four (green) and five (yellow) jets at third order in perturbative QCD for Higgs decay to bottom quarks (left) and gluons (centre). The right frame shows the contribution of the different decay modes to the total hadronic two-jet fraction.}
		\label{fig:rates}
	\end{figure*}
	
	In the last frame in Figure~\ref{fig:3jetNNLO} we compare the impact of the different
	decay modes on the total three-jet decay rate at NNLO, obtained by the weighted sum of the bottom, charm, and gluonic  contributions normalized to~\eqref{eq:gamma_tot}. 
	We observe that the contribution of the gluonic decay mode vanishes for $\ycut\approx 0.001$ and grows to up to $25$\% at $\ycut=0.1$. Hard three-jet final states thus offer the highest 
	sensitivity to the gluonic Higgs boson decay mode. 
		 
	In the first two frames of Figure~\ref{fig:rates}, we present the  jet fractions for different multiplicities at third order in the  QCD coupling. 
	The results expose clear differences between the two decay modes. For the decay to bottom quarks, the jet fractions exhibit similar features to the ones observed in electron-positron annihilation~\cite{Gehrmann-DeRidder:2008qsl}. Indeed, in both cases the underlying Born-level event is given by a colour-singlet decaying to a fermion pair, resulting in an identical structure of QCD corrections.
	On the other hand, the gluonic mode presents significantly larger fractions of higher-multiplicity jet
	final states for any value of $\ycut$. The intersection of the two-jet curve with the three- and four-jet ones occurs for the gluonic decay mode at values of $\ycut$ almost an order of magnitude larger than for the Yukawa-induced mode. 
	
	In the right frame of Figure~\ref{fig:rates}, we study the impact of the different
	decay modes on the total two-jet
	fraction  at N$^3$LO. 
	The contribution of the gluonic decay mode monotonically decreases when smaller values of $\ycut$ are considered. For $\ycut>1/3$ each event is classified as a two-jet event and the the different decay modes respectively stabilise at $84.7$\% (bottom Yukawa), $11.3$\% (gluonic) and $4.0$\% (charm Yukawa) of the total hadronic two-jet rate, in agreement with their contribution to the inclusive hadronic branching ratio~\cite{Baglio:2010ae}. By comparing to the right frame of Figure~\ref{fig:3jetNNLO}, we observe that 
	by selecting hard three-jet final states, the contribution of the gluonic decay mode is doubled compared 
	to a fully exclusive measurement. 
	
\begin{figure}[t]
		\centering
		\includegraphics[width=0.45\textwidth]{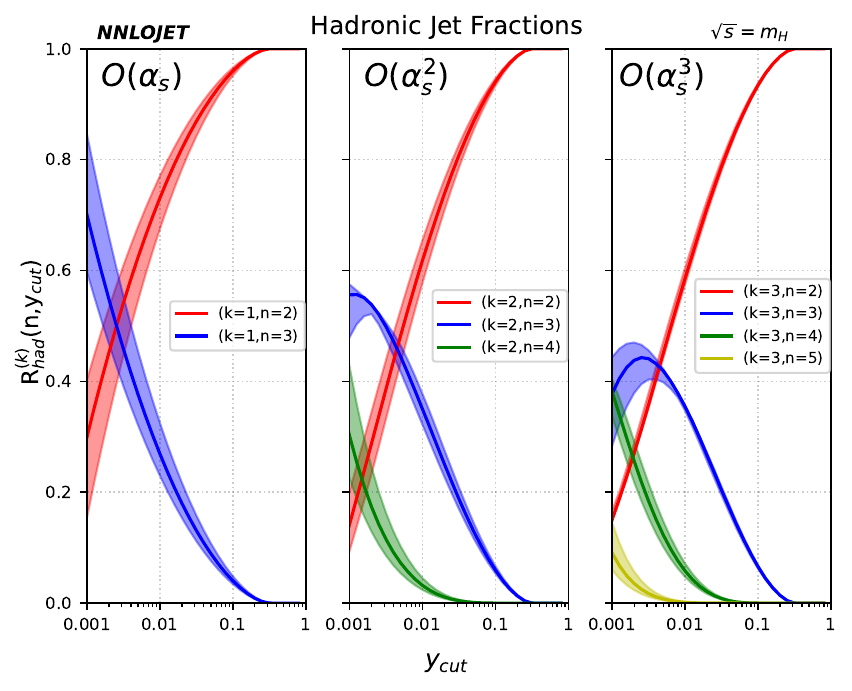}
		\caption{Fractional jet rates for final states containing two (red), three (blue), four (green) and five (yellow) jets in the hadronic decays of a Higgs boson at first (left), second (centre) and third (right) order in perturbative QCD.}
		\label{fig:rates_pert}
	\end{figure}

	In Figure~\ref{fig:rates_pert} we present the two-jet fractions in hadronic decays of the Higgs at first, second, and third order in perturbative QCD.
	The features one observes here are similar to those for the Yukawa-induced mode, that dominates the weighted sum, shown in Figure~\ref{fig:rates} (left).	
At higher orders in perturbation theory, the inclusion of multiple emissions yields non-vanishing four- and five-jet rates. When these open up, the three-jet rates curve ceases to increase at low $\ycut$ value and develops a peak. If even more emissions were considered, one would observe the same behaviour for the four- and five-jet rates as well. Again, to properly capture these effects at low $\ycut$ values, where the sensitivity to multiple emissions is enhanced, all-order resummation will be required.

	In this Letter, we
performed a fully exclusive calculation of the hadronic decay of a Higgs boson to third perturbative order in the strong 
coupling. Our calculation accounts for the two dominant 
types of decay categories: Yukawa-induced to bottom or charm
quarks and heavy-quark-loop-induced to gluons, with the latter not being considered previously at this order. 
Our results allow us to determine Higgs-boson decay rates into three-jet final states to NNLO and into two-jet final states 
to N$^3$LO. Our implementation uses the recently proposed generalised antenna functions~\cite{Fox:2024bfp}, constructed by means of the designer antenna algorithm~\cite{Braun-White:2023sgd,Braun-White:2023zwd}, for final-state radiation, 
representing its first application to a new process. 
 
We observe substantial differences between the two channels: for fixed jet 
resolution $\ycut$ the gluonic mode yields more frequently 
final states with higher jet multiplicity than the Yukawa-induced process. Consequently, three-jet fraction peaks at
larger values of $\ycut$ in the gluonic mode, as expected from the enhancement of radiation patterns off gluons 
compared to quarks. 
Our results indicate substantial differences in the phenomenology of hadronic final states between different 
types of decay categories,
highlighting in particular the enhanced sensitivity on the gluonic decay mode that 
can be obtained by selecting hard three-jet final states, 
and provide accurate QCD predictions to exploit these in future precision 
Higgs studies.	
	\begin{acknowledgments}
		\sect{Acknowledgments}
		We are grateful to Xuan Chen and Ciaran Williams for their respective contributions to the implementation of various Higgs decay matrix elements in \textsc{NNLOjet}~\cite{Chen:2016zka} and \textsc{Eerad3}~\cite{Gehrmann-DeRidder:2023uld}, 
		which were adapted for this work. 
		AG acknowledges the support of the Swiss National Science Foundation (SNF) under contract 200021-231259 and of the Swiss National Supercomputing Centre (CSCS) under project ID ETH5f. 
		TG has received funding from the Swiss National Science Foundation (SNF)
		under contract 200020-204200 and from the European Research Council (ERC) under
		the European Union's Horizon 2020 research and innovation programme grant
		agreement 101019620 (ERC Advanced Grant TOPUP) and acknowledges the hospitality of 
		the Kavli Institute for Theoretical Physics (KITP) Santa Barbara, supported in part by grant NSF PHY-2309135.
		NG gratefully acknowledges support from the UK Science and Technology Facilities Council (STFC) under contract ST/X000745/1 and hospitality from the Pauli Center for Theoretical Studies, Zurich. 
		MM is supported by a Royal Society Newton International Fellowship (NIF/R1/232539).
		Part of the computations were carried out on the PLEIADES cluster at the University of Wuppertal, supported by the Deutsche Forschungsgemeinschaft (DFG, grant No. INST 218/78-1 FUGG) and the Bundesministerium f\"{u}r Bildung und Forschung (BMBF).
	\end{acknowledgments}

	\bibliography{letter}

\begin{thebibliography}{51}%
\makeatletter
\providecommand \@ifxundefined [1]{%
 \@ifx{#1\undefined}
}%
\providecommand \@ifnum [1]{%
 \ifnum #1\expandafter \@firstoftwo
 \else \expandafter \@secondoftwo
 \fi
}%
\providecommand \@ifx [1]{%
 \ifx #1\expandafter \@firstoftwo
 \else \expandafter \@secondoftwo
 \fi
}%
\providecommand \natexlab [1]{#1}%
\providecommand \enquote  [1]{``#1''}%
\providecommand \bibnamefont  [1]{#1}%
\providecommand \bibfnamefont [1]{#1}%
\providecommand \citenamefont [1]{#1}%
\providecommand \href@noop [0]{\@secondoftwo}%
\providecommand \href [0]{\begingroup \@sanitize@url \@href}%
\providecommand \@href[1]{\@@startlink{#1}\@@href}%
\providecommand \@@href[1]{\endgroup#1\@@endlink}%
\providecommand \@sanitize@url [0]{\catcode `\\12\catcode `\$12\catcode
  `\&12\catcode `\#12\catcode `\^12\catcode `\_12\catcode `\%12\relax}%
\providecommand \@@startlink[1]{}%
\providecommand \@@endlink[0]{}%
\providecommand \url  [0]{\begingroup\@sanitize@url \@url }%
\providecommand \@url [1]{\endgroup\@href {#1}{\urlprefix }}%
\providecommand \urlprefix  [0]{URL }%
\providecommand \Eprint [0]{\href }%
\providecommand \doibase [0]{http://dx.doi.org/}%
\providecommand \selectlanguage [0]{\@gobble}%
\providecommand \bibinfo  [0]{\@secondoftwo}%
\providecommand \bibfield  [0]{\@secondoftwo}%
\providecommand \translation [1]{[#1]}%
\providecommand \BibitemOpen [0]{}%
\providecommand \bibitemStop [0]{}%
\providecommand \bibitemNoStop [0]{.\EOS\space}%
\providecommand \EOS [0]{\spacefactor3000\relax}%
\providecommand \BibitemShut  [1]{\csname bibitem#1\endcsname}%
\let\auto@bib@innerbib\@empty
\bibitem [{\citenamefont {Aad}\ \emph {et~al.}(2012)\citenamefont {Aad} \emph
  {et~al.}}]{ATLAS:2012yve}%
  \BibitemOpen
  \bibfield  {author} {\bibinfo {author} {\bibfnamefont {G.}~\bibnamefont
  {Aad}} \emph {et~al.} (\bibinfo {collaboration} {ATLAS}),\ }\href {\doibase
  10.1016/j.physletb.2012.08.020} {\bibfield  {journal} {\bibinfo  {journal}
  {Phys. Lett. B}\ }\textbf {\bibinfo {volume} {716}},\ \bibinfo {pages} {1}
  (\bibinfo {year} {2012})},\ \Eprint {http://arxiv.org/abs/1207.7214}
  {arXiv:1207.7214 [hep-ex]} \BibitemShut {NoStop}%
\bibitem [{\citenamefont {Chatrchyan}\ \emph {et~al.}(2012)\citenamefont
  {Chatrchyan} \emph {et~al.}}]{CMS:2012qbp}%
  \BibitemOpen
  \bibfield  {author} {\bibinfo {author} {\bibfnamefont {S.}~\bibnamefont
  {Chatrchyan}} \emph {et~al.} (\bibinfo {collaboration} {CMS}),\ }\href
  {\doibase 10.1016/j.physletb.2012.08.021} {\bibfield  {journal} {\bibinfo
  {journal} {Phys. Lett. B}\ }\textbf {\bibinfo {volume} {716}},\ \bibinfo
  {pages} {30} (\bibinfo {year} {2012})},\ \Eprint
  {http://arxiv.org/abs/1207.7235} {arXiv:1207.7235 [hep-ex]} \BibitemShut
  {NoStop}%
\bibitem [{\citenamefont {Aad}\ \emph {et~al.}(2022)\citenamefont {Aad} \emph
  {et~al.}}]{ATLAS:2022vkf}%
  \BibitemOpen
  \bibfield  {author} {\bibinfo {author} {\bibfnamefont {G.}~\bibnamefont
  {Aad}} \emph {et~al.} (\bibinfo {collaboration} {ATLAS}),\ }\href {\doibase
  10.1038/s41586-022-04893-w} {\bibfield  {journal} {\bibinfo  {journal}
  {Nature}\ }\textbf {\bibinfo {volume} {607}},\ \bibinfo {pages} {52}
  (\bibinfo {year} {2022})},\ \bibinfo {note} {[Erratum: Nature 612, E24
  (2022)]},\ \Eprint {http://arxiv.org/abs/2207.00092} {arXiv:2207.00092
  [hep-ex]} \BibitemShut {NoStop}%
\bibitem [{\citenamefont {Tumasyan}\ \emph {et~al.}(2022)\citenamefont
  {Tumasyan} \emph {et~al.}}]{CMS:2022dwd}%
  \BibitemOpen
  \bibfield  {author} {\bibinfo {author} {\bibfnamefont {A.}~\bibnamefont
  {Tumasyan}} \emph {et~al.} (\bibinfo {collaboration} {CMS}),\ }\href
  {\doibase 10.1038/s41586-022-04892-x} {\bibfield  {journal} {\bibinfo
  {journal} {Nature}\ }\textbf {\bibinfo {volume} {607}},\ \bibinfo {pages}
  {60} (\bibinfo {year} {2022})},\ \bibinfo {note} {[Erratum: Nature 623,
  (2023)]},\ \Eprint {http://arxiv.org/abs/2207.00043} {arXiv:2207.00043
  [hep-ex]} \BibitemShut {NoStop}%
\bibitem [{\citenamefont {Aaboud}\ \emph {et~al.}(2018)\citenamefont {Aaboud}
  \emph {et~al.}}]{ATLAS:2018kot}%
  \BibitemOpen
  \bibfield  {author} {\bibinfo {author} {\bibfnamefont {M.}~\bibnamefont
  {Aaboud}} \emph {et~al.} (\bibinfo {collaboration} {ATLAS}),\ }\href
  {\doibase 10.1016/j.physletb.2018.09.013} {\bibfield  {journal} {\bibinfo
  {journal} {Phys. Lett. B}\ }\textbf {\bibinfo {volume} {786}},\ \bibinfo
  {pages} {59} (\bibinfo {year} {2018})},\ \Eprint
  {http://arxiv.org/abs/1808.08238} {arXiv:1808.08238 [hep-ex]} \BibitemShut
  {NoStop}%
\bibitem [{\citenamefont {Sirunyan}\ \emph {et~al.}(2018)\citenamefont
  {Sirunyan} \emph {et~al.}}]{CMS:2018nsn}%
  \BibitemOpen
  \bibfield  {author} {\bibinfo {author} {\bibfnamefont {A.~M.}\ \bibnamefont
  {Sirunyan}} \emph {et~al.} (\bibinfo {collaboration} {CMS}),\ }\href
  {\doibase 10.1103/PhysRevLett.121.121801} {\bibfield  {journal} {\bibinfo
  {journal} {Phys. Rev. Lett.}\ }\textbf {\bibinfo {volume} {121}},\ \bibinfo
  {pages} {121801} (\bibinfo {year} {2018})},\ \Eprint
  {http://arxiv.org/abs/1808.08242} {arXiv:1808.08242 [hep-ex]} \BibitemShut
  {NoStop}%
\bibitem [{\citenamefont {Abada}\ \emph
  {et~al.}(2019{\natexlab{a}})\citenamefont {Abada} \emph
  {et~al.}}]{FCC:2018byv}%
  \BibitemOpen
  \bibfield  {author} {\bibinfo {author} {\bibfnamefont {A.}~\bibnamefont
  {Abada}} \emph {et~al.} (\bibinfo {collaboration} {FCC}),\ }\href {\doibase
  10.1140/epjc/s10052-019-6904-3} {\bibfield  {journal} {\bibinfo  {journal}
  {Eur. Phys. J. C}\ }\textbf {\bibinfo {volume} {79}},\ \bibinfo {pages} {474}
  (\bibinfo {year} {2019}{\natexlab{a}})}\BibitemShut {NoStop}%
\bibitem [{\citenamefont {Abada}\ \emph
  {et~al.}(2019{\natexlab{b}})\citenamefont {Abada} \emph
  {et~al.}}]{FCC:2018evy}%
  \BibitemOpen
  \bibfield  {author} {\bibinfo {author} {\bibfnamefont {A.}~\bibnamefont
  {Abada}} \emph {et~al.} (\bibinfo {collaboration} {FCC}),\ }\href {\doibase
  10.1140/epjst/e2019-900045-4} {\bibfield  {journal} {\bibinfo  {journal}
  {Eur. Phys. J. ST}\ }\textbf {\bibinfo {volume} {228}},\ \bibinfo {pages}
  {261} (\bibinfo {year} {2019}{\natexlab{b}})}\BibitemShut {NoStop}%
\bibitem [{\citenamefont {Dong}\ \emph {et~al.}(2018)\citenamefont {Dong} \emph
  {et~al.}}]{CEPCStudyGroup:2018ghi}%
  \BibitemOpen
  \bibfield  {author} {\bibinfo {author} {\bibfnamefont {M.}~\bibnamefont
  {Dong}} \emph {et~al.} (\bibinfo {collaboration} {CEPC Study Group}),\
  }\href@noop {} {\  (\bibinfo {year} {2018})},\ \Eprint
  {http://arxiv.org/abs/1811.10545} {arXiv:1811.10545 [hep-ex]} \BibitemShut
  {NoStop}%
\bibitem [{\citenamefont {Baikov}\ \emph {et~al.}(2006)\citenamefont {Baikov},
  \citenamefont {Chetyrkin},\ and\ \citenamefont {Kuhn}}]{Baikov:2005rw}%
  \BibitemOpen
  \bibfield  {author} {\bibinfo {author} {\bibfnamefont {P.~A.}\ \bibnamefont
  {Baikov}}, \bibinfo {author} {\bibfnamefont {K.~G.}\ \bibnamefont
  {Chetyrkin}}, \ and\ \bibinfo {author} {\bibfnamefont {J.~H.}\ \bibnamefont
  {Kuhn}},\ }\href {\doibase 10.1103/PhysRevLett.96.012003} {\bibfield
  {journal} {\bibinfo  {journal} {Phys. Rev. Lett.}\ }\textbf {\bibinfo
  {volume} {96}},\ \bibinfo {pages} {012003} (\bibinfo {year} {2006})},\
  \Eprint {http://arxiv.org/abs/hep-ph/0511063} {arXiv:hep-ph/0511063}
  \BibitemShut {NoStop}%
\bibitem [{\citenamefont {Baikov}\ and\ \citenamefont
  {Chetyrkin}(2006)}]{Baikov:2006ch}%
  \BibitemOpen
  \bibfield  {author} {\bibinfo {author} {\bibfnamefont {P.~A.}\ \bibnamefont
  {Baikov}}\ and\ \bibinfo {author} {\bibfnamefont {K.~G.}\ \bibnamefont
  {Chetyrkin}},\ }\href {\doibase 10.1103/PhysRevLett.97.061803} {\bibfield
  {journal} {\bibinfo  {journal} {Phys. Rev. Lett.}\ }\textbf {\bibinfo
  {volume} {97}},\ \bibinfo {pages} {061803} (\bibinfo {year} {2006})},\
  \Eprint {http://arxiv.org/abs/hep-ph/0604194} {arXiv:hep-ph/0604194}
  \BibitemShut {NoStop}%
\bibitem [{\citenamefont {Davies}\ \emph {et~al.}(2017)\citenamefont {Davies},
  \citenamefont {Steinhauser},\ and\ \citenamefont
  {Wellmann}}]{Davies:2017xsp}%
  \BibitemOpen
  \bibfield  {author} {\bibinfo {author} {\bibfnamefont {J.}~\bibnamefont
  {Davies}}, \bibinfo {author} {\bibfnamefont {M.}~\bibnamefont {Steinhauser}},
  \ and\ \bibinfo {author} {\bibfnamefont {D.}~\bibnamefont {Wellmann}},\
  }\href {\doibase 10.1016/j.nuclphysb.2017.04.012} {\bibfield  {journal}
  {\bibinfo  {journal} {Nucl. Phys. B}\ }\textbf {\bibinfo {volume} {920}},\
  \bibinfo {pages} {20} (\bibinfo {year} {2017})},\ \Eprint
  {http://arxiv.org/abs/1703.02988} {arXiv:1703.02988 [hep-ph]} \BibitemShut
  {NoStop}%
\bibitem [{\citenamefont {Herzog}\ \emph {et~al.}(2017)\citenamefont {Herzog},
  \citenamefont {Ruijl}, \citenamefont {Ueda}, \citenamefont {Vermaseren},\
  and\ \citenamefont {Vogt}}]{Herzog:2017dtz}%
  \BibitemOpen
  \bibfield  {author} {\bibinfo {author} {\bibfnamefont {F.}~\bibnamefont
  {Herzog}}, \bibinfo {author} {\bibfnamefont {B.}~\bibnamefont {Ruijl}},
  \bibinfo {author} {\bibfnamefont {T.}~\bibnamefont {Ueda}}, \bibinfo {author}
  {\bibfnamefont {J.~A.~M.}\ \bibnamefont {Vermaseren}}, \ and\ \bibinfo
  {author} {\bibfnamefont {A.}~\bibnamefont {Vogt}},\ }\href {\doibase
  10.1007/JHEP08(2017)113} {\bibfield  {journal} {\bibinfo  {journal} {JHEP}\
  }\textbf {\bibinfo {volume} {08}},\ \bibinfo {pages} {113} (\bibinfo {year}
  {2017})},\ \Eprint {http://arxiv.org/abs/1707.01044} {arXiv:1707.01044
  [hep-ph]} \BibitemShut {NoStop}%
\bibitem [{\citenamefont {Mondini}\ and\ \citenamefont
  {Williams}(2019)}]{Mondini:2019vub}%
  \BibitemOpen
  \bibfield  {author} {\bibinfo {author} {\bibfnamefont {R.}~\bibnamefont
  {Mondini}}\ and\ \bibinfo {author} {\bibfnamefont {C.}~\bibnamefont
  {Williams}},\ }\href {\doibase 10.1007/JHEP06(2019)120} {\bibfield  {journal}
  {\bibinfo  {journal} {JHEP}\ }\textbf {\bibinfo {volume} {06}},\ \bibinfo
  {pages} {120} (\bibinfo {year} {2019})},\ \Eprint
  {http://arxiv.org/abs/1904.08961} {arXiv:1904.08961 [hep-ph]} \BibitemShut
  {NoStop}%
\bibitem [{\citenamefont {Mondini}\ \emph {et~al.}(2019)\citenamefont
  {Mondini}, \citenamefont {Schiavi},\ and\ \citenamefont
  {Williams}}]{Mondini:2019gid}%
  \BibitemOpen
  \bibfield  {author} {\bibinfo {author} {\bibfnamefont {R.}~\bibnamefont
  {Mondini}}, \bibinfo {author} {\bibfnamefont {M.}~\bibnamefont {Schiavi}}, \
  and\ \bibinfo {author} {\bibfnamefont {C.}~\bibnamefont {Williams}},\ }\href
  {\doibase 10.1007/JHEP06(2019)079} {\bibfield  {journal} {\bibinfo  {journal}
  {JHEP}\ }\textbf {\bibinfo {volume} {06}},\ \bibinfo {pages} {079} (\bibinfo
  {year} {2019})},\ \Eprint {http://arxiv.org/abs/1904.08960} {arXiv:1904.08960
  [hep-ph]} \BibitemShut {NoStop}%
\bibitem [{\citenamefont {Bernreuther}\ \emph {et~al.}(2018)\citenamefont
  {Bernreuther}, \citenamefont {Chen},\ and\ \citenamefont
  {Si}}]{Bernreuther:2018ynm}%
  \BibitemOpen
  \bibfield  {author} {\bibinfo {author} {\bibfnamefont {W.}~\bibnamefont
  {Bernreuther}}, \bibinfo {author} {\bibfnamefont {L.}~\bibnamefont {Chen}}, \
  and\ \bibinfo {author} {\bibfnamefont {Z.-G.}\ \bibnamefont {Si}},\ }\href
  {\doibase 10.1007/JHEP07(2018)159} {\bibfield  {journal} {\bibinfo  {journal}
  {JHEP}\ }\textbf {\bibinfo {volume} {07}},\ \bibinfo {pages} {159} (\bibinfo
  {year} {2018})},\ \Eprint {http://arxiv.org/abs/1805.06658} {arXiv:1805.06658
  [hep-ph]} \BibitemShut {NoStop}%
\bibitem [{\citenamefont {Coloretti}\ \emph {et~al.}(2022)\citenamefont
  {Coloretti}, \citenamefont {Gehrmann-De~Ridder},\ and\ \citenamefont
  {Preuss}}]{Coloretti:2022jcl}%
  \BibitemOpen
  \bibfield  {author} {\bibinfo {author} {\bibfnamefont {G.}~\bibnamefont
  {Coloretti}}, \bibinfo {author} {\bibfnamefont {A.}~\bibnamefont
  {Gehrmann-De~Ridder}}, \ and\ \bibinfo {author} {\bibfnamefont {C.~T.}\
  \bibnamefont {Preuss}},\ }\href {\doibase 10.1007/JHEP06(2022)009} {\bibfield
   {journal} {\bibinfo  {journal} {JHEP}\ }\textbf {\bibinfo {volume} {06}},\
  \bibinfo {pages} {009} (\bibinfo {year} {2022})},\ \Eprint
  {http://arxiv.org/abs/2202.07333} {arXiv:2202.07333 [hep-ph]} \BibitemShut
  {NoStop}%
\bibitem [{\citenamefont {Gehrmann-De~Ridder}\ \emph
  {et~al.}(2024{\natexlab{a}})\citenamefont {Gehrmann-De~Ridder}, \citenamefont
  {Preuss}, \citenamefont {Reichelt},\ and\ \citenamefont
  {Schumann}}]{Gehrmann-DeRidder:2024avt}%
  \BibitemOpen
  \bibfield  {author} {\bibinfo {author} {\bibfnamefont {A.}~\bibnamefont
  {Gehrmann-De~Ridder}}, \bibinfo {author} {\bibfnamefont {C.~T.}\ \bibnamefont
  {Preuss}}, \bibinfo {author} {\bibfnamefont {D.}~\bibnamefont {Reichelt}}, \
  and\ \bibinfo {author} {\bibfnamefont {S.}~\bibnamefont {Schumann}},\ }\href
  {\doibase 10.1007/JHEP07(2024)160} {\bibfield  {journal} {\bibinfo  {journal}
  {JHEP}\ }\textbf {\bibinfo {volume} {07}},\ \bibinfo {pages} {160} (\bibinfo
  {year} {2024}{\natexlab{a}})},\ \Eprint {http://arxiv.org/abs/2403.06929}
  {arXiv:2403.06929 [hep-ph]} \BibitemShut {NoStop}%
\bibitem [{\citenamefont {Gehrmann-De~Ridder}\ \emph
  {et~al.}(2024{\natexlab{b}})\citenamefont {Gehrmann-De~Ridder}, \citenamefont
  {Preuss},\ and\ \citenamefont {Williams}}]{Gehrmann-DeRidder:2023uld}%
  \BibitemOpen
  \bibfield  {author} {\bibinfo {author} {\bibfnamefont {A.}~\bibnamefont
  {Gehrmann-De~Ridder}}, \bibinfo {author} {\bibfnamefont {C.~T.}\ \bibnamefont
  {Preuss}}, \ and\ \bibinfo {author} {\bibfnamefont {C.}~\bibnamefont
  {Williams}},\ }\href {\doibase 10.1007/JHEP03(2024)104} {\bibfield  {journal}
  {\bibinfo  {journal} {JHEP}\ }\textbf {\bibinfo {volume} {03}},\ \bibinfo
  {pages} {104} (\bibinfo {year} {2024}{\natexlab{b}})},\ \Eprint
  {http://arxiv.org/abs/2310.09354} {arXiv:2310.09354 [hep-ph]} \BibitemShut
  {NoStop}%
\bibitem [{\citenamefont {Gao}(2018)}]{Gao:2016jcm}%
  \BibitemOpen
  \bibfield  {author} {\bibinfo {author} {\bibfnamefont {J.}~\bibnamefont
  {Gao}},\ }\href {\doibase 10.1007/JHEP01(2018)038} {\bibfield  {journal}
  {\bibinfo  {journal} {JHEP}\ }\textbf {\bibinfo {volume} {01}},\ \bibinfo
  {pages} {038} (\bibinfo {year} {2018})},\ \Eprint
  {http://arxiv.org/abs/1608.01746} {arXiv:1608.01746 [hep-ph]} \BibitemShut
  {NoStop}%
\bibitem [{\citenamefont {Knobbe}\ \emph {et~al.}(2024)\citenamefont {Knobbe},
  \citenamefont {Krauss}, \citenamefont {Reichelt},\ and\ \citenamefont
  {Schumann}}]{Knobbe:2023njd}%
  \BibitemOpen
  \bibfield  {author} {\bibinfo {author} {\bibfnamefont {M.}~\bibnamefont
  {Knobbe}}, \bibinfo {author} {\bibfnamefont {F.}~\bibnamefont {Krauss}},
  \bibinfo {author} {\bibfnamefont {D.}~\bibnamefont {Reichelt}}, \ and\
  \bibinfo {author} {\bibfnamefont {S.}~\bibnamefont {Schumann}},\ }\href
  {\doibase 10.1140/epjc/s10052-024-12430-4} {\bibfield  {journal} {\bibinfo
  {journal} {Eur. Phys. J. C}\ }\textbf {\bibinfo {volume} {84}},\ \bibinfo
  {pages} {83} (\bibinfo {year} {2024})},\ \Eprint
  {http://arxiv.org/abs/2306.03682} {arXiv:2306.03682 [hep-ph]} \BibitemShut
  {NoStop}%
\bibitem [{\citenamefont {Campillo~Aveleira}\ \emph {et~al.}(2024)\citenamefont
  {Campillo~Aveleira}, \citenamefont {Gehrmann-De~Ridder},\ and\ \citenamefont
  {Preuss}}]{CampilloAveleira:2024fll}%
  \BibitemOpen
  \bibfield  {author} {\bibinfo {author} {\bibfnamefont {B.}~\bibnamefont
  {Campillo~Aveleira}}, \bibinfo {author} {\bibfnamefont {A.}~\bibnamefont
  {Gehrmann-De~Ridder}}, \ and\ \bibinfo {author} {\bibfnamefont {C.~T.}\
  \bibnamefont {Preuss}},\ }\href {\doibase 10.1140/epjc/s10052-024-13127-4}
  {\bibfield  {journal} {\bibinfo  {journal} {Eur. Phys. J. C}\ }\textbf
  {\bibinfo {volume} {84}},\ \bibinfo {pages} {789} (\bibinfo {year} {2024})},\
  \Eprint {http://arxiv.org/abs/2402.17379} {arXiv:2402.17379 [hep-ph]}
  \BibitemShut {NoStop}%
\bibitem [{\citenamefont {Wilczek}(1977)}]{Wilczek:1977zn}%
  \BibitemOpen
  \bibfield  {author} {\bibinfo {author} {\bibfnamefont {F.}~\bibnamefont
  {Wilczek}},\ }\href {\doibase 10.1103/PhysRevLett.39.1304} {\bibfield
  {journal} {\bibinfo  {journal} {Phys. Rev. Lett.}\ }\textbf {\bibinfo
  {volume} {39}},\ \bibinfo {pages} {1304} (\bibinfo {year}
  {1977})}\BibitemShut {NoStop}%
\bibitem [{\citenamefont {Shifman}\ \emph {et~al.}(1978)\citenamefont
  {Shifman}, \citenamefont {Vainshtein},\ and\ \citenamefont
  {Zakharov}}]{Shifman:1978zn}%
  \BibitemOpen
  \bibfield  {author} {\bibinfo {author} {\bibfnamefont {M.~A.}\ \bibnamefont
  {Shifman}}, \bibinfo {author} {\bibfnamefont {A.~I.}\ \bibnamefont
  {Vainshtein}}, \ and\ \bibinfo {author} {\bibfnamefont {V.~I.}\ \bibnamefont
  {Zakharov}},\ }\href {\doibase 10.1016/0370-2693(78)90481-1} {\bibfield
  {journal} {\bibinfo  {journal} {Phys. Lett. B}\ }\textbf {\bibinfo {volume}
  {78}},\ \bibinfo {pages} {443} (\bibinfo {year} {1978})}\BibitemShut
  {NoStop}%
\bibitem [{\citenamefont {Inami}\ \emph {et~al.}(1983)\citenamefont {Inami},
  \citenamefont {Kubota},\ and\ \citenamefont {Okada}}]{Inami:1982xt}%
  \BibitemOpen
  \bibfield  {author} {\bibinfo {author} {\bibfnamefont {T.}~\bibnamefont
  {Inami}}, \bibinfo {author} {\bibfnamefont {T.}~\bibnamefont {Kubota}}, \
  and\ \bibinfo {author} {\bibfnamefont {Y.}~\bibnamefont {Okada}},\ }\href
  {\doibase 10.1007/BF01571710} {\bibfield  {journal} {\bibinfo  {journal} {Z.
  Phys. C}\ }\textbf {\bibinfo {volume} {18}},\ \bibinfo {pages} {69} (\bibinfo
  {year} {1983})}\BibitemShut {NoStop}%
\bibitem [{\citenamefont {Catani}\ \emph {et~al.}(1991)\citenamefont {Catani},
  \citenamefont {Dokshitzer}, \citenamefont {Olsson}, \citenamefont {Turnock},\
  and\ \citenamefont {Webber}}]{Catani:1991hj}%
  \BibitemOpen
  \bibfield  {author} {\bibinfo {author} {\bibfnamefont {S.}~\bibnamefont
  {Catani}}, \bibinfo {author} {\bibfnamefont {Y.~L.}\ \bibnamefont
  {Dokshitzer}}, \bibinfo {author} {\bibfnamefont {M.}~\bibnamefont {Olsson}},
  \bibinfo {author} {\bibfnamefont {G.}~\bibnamefont {Turnock}}, \ and\
  \bibinfo {author} {\bibfnamefont {B.~R.}\ \bibnamefont {Webber}},\ }\href
  {\doibase 10.1016/0370-2693(91)90196-W} {\bibfield  {journal} {\bibinfo
  {journal} {Phys. Lett. B}\ }\textbf {\bibinfo {volume} {269}},\ \bibinfo
  {pages} {432} (\bibinfo {year} {1991})}\BibitemShut {NoStop}%
\bibitem [{\citenamefont {Brown}\ and\ \citenamefont
  {Stirling}(1990)}]{Brown:1990nm}%
  \BibitemOpen
  \bibfield  {author} {\bibinfo {author} {\bibfnamefont {N.}~\bibnamefont
  {Brown}}\ and\ \bibinfo {author} {\bibfnamefont {W.~J.}\ \bibnamefont
  {Stirling}},\ }\href {\doibase 10.1016/0370-2693(90)90502-W} {\bibfield
  {journal} {\bibinfo  {journal} {Phys. Lett. B}\ }\textbf {\bibinfo {volume}
  {252}},\ \bibinfo {pages} {657} (\bibinfo {year} {1990})}\BibitemShut
  {NoStop}%
\bibitem [{\citenamefont {Brown}\ and\ \citenamefont
  {Stirling}(1992)}]{Brown:1991hx}%
  \BibitemOpen
  \bibfield  {author} {\bibinfo {author} {\bibfnamefont {N.}~\bibnamefont
  {Brown}}\ and\ \bibinfo {author} {\bibfnamefont {W.~J.}\ \bibnamefont
  {Stirling}},\ }\href {\doibase 10.1007/BF01559740} {\bibfield  {journal}
  {\bibinfo  {journal} {Z. Phys. C}\ }\textbf {\bibinfo {volume} {53}},\
  \bibinfo {pages} {629} (\bibinfo {year} {1992})}\BibitemShut {NoStop}%
\bibitem [{\citenamefont {Stirling}(1991)}]{Stirling:1991ds}%
  \BibitemOpen
  \bibfield  {author} {\bibinfo {author} {\bibfnamefont {W.~J.}\ \bibnamefont
  {Stirling}},\ }\href {\doibase 10.1088/0954-3899/17/10/014} {\bibfield
  {journal} {\bibinfo  {journal} {J. Phys. G}\ }\textbf {\bibinfo {volume}
  {17}},\ \bibinfo {pages} {1567} (\bibinfo {year} {1991})}\BibitemShut
  {NoStop}%
\bibitem [{\citenamefont {Bethke}\ \emph {et~al.}(1992)\citenamefont {Bethke},
  \citenamefont {Kunszt}, \citenamefont {Soper},\ and\ \citenamefont
  {Stirling}}]{Bethke:1991wk}%
  \BibitemOpen
  \bibfield  {author} {\bibinfo {author} {\bibfnamefont {S.}~\bibnamefont
  {Bethke}}, \bibinfo {author} {\bibfnamefont {Z.}~\bibnamefont {Kunszt}},
  \bibinfo {author} {\bibfnamefont {D.~E.}\ \bibnamefont {Soper}}, \ and\
  \bibinfo {author} {\bibfnamefont {W.~J.}\ \bibnamefont {Stirling}},\ }\href
  {\doibase 10.1016/0550-3213(92)90289-N} {\bibfield  {journal} {\bibinfo
  {journal} {Nucl. Phys. B}\ }\textbf {\bibinfo {volume} {370}},\ \bibinfo
  {pages} {310} (\bibinfo {year} {1992})},\ \bibinfo {note} {[Erratum:
  Nucl.Phys.B 523, 681--681 (1998)]}\BibitemShut {NoStop}%
\bibitem [{\citenamefont {Gehrmann-De~Ridder}\ \emph
  {et~al.}(2008)\citenamefont {Gehrmann-De~Ridder}, \citenamefont {Gehrmann},
  \citenamefont {Glover},\ and\ \citenamefont
  {Heinrich}}]{Gehrmann-DeRidder:2008qsl}%
  \BibitemOpen
  \bibfield  {author} {\bibinfo {author} {\bibfnamefont {A.}~\bibnamefont
  {Gehrmann-De~Ridder}}, \bibinfo {author} {\bibfnamefont {T.}~\bibnamefont
  {Gehrmann}}, \bibinfo {author} {\bibfnamefont {E.~W.~N.}\ \bibnamefont
  {Glover}}, \ and\ \bibinfo {author} {\bibfnamefont {G.}~\bibnamefont
  {Heinrich}},\ }\href {\doibase 10.1103/PhysRevLett.100.172001} {\bibfield
  {journal} {\bibinfo  {journal} {Phys. Rev. Lett.}\ }\textbf {\bibinfo
  {volume} {100}},\ \bibinfo {pages} {172001} (\bibinfo {year} {2008})},\
  \Eprint {http://arxiv.org/abs/0802.0813} {arXiv:0802.0813 [hep-ph]}
  \BibitemShut {NoStop}%
\bibitem [{\citenamefont {Del~Duca}\ \emph {et~al.}(2004)\citenamefont
  {Del~Duca}, \citenamefont {Frizzo},\ and\ \citenamefont
  {Maltoni}}]{DelDuca:2004wt}%
  \BibitemOpen
  \bibfield  {author} {\bibinfo {author} {\bibfnamefont {V.}~\bibnamefont
  {Del~Duca}}, \bibinfo {author} {\bibfnamefont {A.}~\bibnamefont {Frizzo}}, \
  and\ \bibinfo {author} {\bibfnamefont {F.}~\bibnamefont {Maltoni}},\ }\href
  {\doibase 10.1088/1126-6708/2004/05/064} {\bibfield  {journal} {\bibinfo
  {journal} {JHEP}\ }\textbf {\bibinfo {volume} {05}},\ \bibinfo {pages} {064}
  (\bibinfo {year} {2004})},\ \Eprint {http://arxiv.org/abs/hep-ph/0404013}
  {arXiv:hep-ph/0404013} \BibitemShut {NoStop}%
\bibitem [{\citenamefont {Anastasiou}\ \emph {et~al.}(2012)\citenamefont
  {Anastasiou}, \citenamefont {Herzog},\ and\ \citenamefont
  {Lazopoulos}}]{Anastasiou:2011qx}%
  \BibitemOpen
  \bibfield  {author} {\bibinfo {author} {\bibfnamefont {C.}~\bibnamefont
  {Anastasiou}}, \bibinfo {author} {\bibfnamefont {F.}~\bibnamefont {Herzog}},
  \ and\ \bibinfo {author} {\bibfnamefont {A.}~\bibnamefont {Lazopoulos}},\
  }\href {\doibase 10.1007/JHEP03(2012)035} {\bibfield  {journal} {\bibinfo
  {journal} {JHEP}\ }\textbf {\bibinfo {volume} {03}},\ \bibinfo {pages} {035}
  (\bibinfo {year} {2012})},\ \Eprint {http://arxiv.org/abs/1110.2368}
  {arXiv:1110.2368 [hep-ph]} \BibitemShut {NoStop}%
\bibitem [{\citenamefont {Del~Duca}\ \emph {et~al.}(2015)\citenamefont
  {Del~Duca}, \citenamefont {Duhr}, \citenamefont {Somogyi}, \citenamefont
  {Tramontano},\ and\ \citenamefont {Tr\'ocs\'anyi}}]{DelDuca:2015zqa}%
  \BibitemOpen
  \bibfield  {author} {\bibinfo {author} {\bibfnamefont {V.}~\bibnamefont
  {Del~Duca}}, \bibinfo {author} {\bibfnamefont {C.}~\bibnamefont {Duhr}},
  \bibinfo {author} {\bibfnamefont {G.}~\bibnamefont {Somogyi}}, \bibinfo
  {author} {\bibfnamefont {F.}~\bibnamefont {Tramontano}}, \ and\ \bibinfo
  {author} {\bibfnamefont {Z.}~\bibnamefont {Tr\'ocs\'anyi}},\ }\href {\doibase
  10.1007/JHEP04(2015)036} {\bibfield  {journal} {\bibinfo  {journal} {JHEP}\
  }\textbf {\bibinfo {volume} {04}},\ \bibinfo {pages} {036} (\bibinfo {year}
  {2015})},\ \Eprint {http://arxiv.org/abs/1501.07226} {arXiv:1501.07226
  [hep-ph]} \BibitemShut {NoStop}%
\bibitem [{\citenamefont {Dixon}\ and\ \citenamefont
  {Sofianatos}(2009)}]{Dixon:2009uk}%
  \BibitemOpen
  \bibfield  {author} {\bibinfo {author} {\bibfnamefont {L.~J.}\ \bibnamefont
  {Dixon}}\ and\ \bibinfo {author} {\bibfnamefont {Y.}~\bibnamefont
  {Sofianatos}},\ }\href {\doibase 10.1088/1126-6708/2009/08/058} {\bibfield
  {journal} {\bibinfo  {journal} {JHEP}\ }\textbf {\bibinfo {volume} {08}},\
  \bibinfo {pages} {058} (\bibinfo {year} {2009})},\ \Eprint
  {http://arxiv.org/abs/0906.0008} {arXiv:0906.0008 [hep-ph]} \BibitemShut
  {NoStop}%
\bibitem [{\citenamefont {Badger}\ \emph {et~al.}(2010)\citenamefont {Badger},
  \citenamefont {Nigel~Glover}, \citenamefont {Mastrolia},\ and\ \citenamefont
  {Williams}}]{Badger:2009hw}%
  \BibitemOpen
  \bibfield  {author} {\bibinfo {author} {\bibfnamefont {S.}~\bibnamefont
  {Badger}}, \bibinfo {author} {\bibfnamefont {E.~W.}\ \bibnamefont
  {Nigel~Glover}}, \bibinfo {author} {\bibfnamefont {P.}~\bibnamefont
  {Mastrolia}}, \ and\ \bibinfo {author} {\bibfnamefont {C.}~\bibnamefont
  {Williams}},\ }\href {\doibase 10.1007/JHEP01(2010)036} {\bibfield  {journal}
  {\bibinfo  {journal} {JHEP}\ }\textbf {\bibinfo {volume} {01}},\ \bibinfo
  {pages} {036} (\bibinfo {year} {2010})},\ \Eprint
  {http://arxiv.org/abs/0909.4475} {arXiv:0909.4475 [hep-ph]} \BibitemShut
  {NoStop}%
\bibitem [{\citenamefont {Badger}\ \emph {et~al.}(2009)\citenamefont {Badger},
  \citenamefont {Campbell}, \citenamefont {Ellis},\ and\ \citenamefont
  {Williams}}]{Badger:2009vh}%
  \BibitemOpen
  \bibfield  {author} {\bibinfo {author} {\bibfnamefont {S.}~\bibnamefont
  {Badger}}, \bibinfo {author} {\bibfnamefont {J.~M.}\ \bibnamefont
  {Campbell}}, \bibinfo {author} {\bibfnamefont {R.~K.}\ \bibnamefont {Ellis}},
  \ and\ \bibinfo {author} {\bibfnamefont {C.}~\bibnamefont {Williams}},\
  }\href {\doibase 10.1088/1126-6708/2009/12/035} {\bibfield  {journal}
  {\bibinfo  {journal} {JHEP}\ }\textbf {\bibinfo {volume} {12}},\ \bibinfo
  {pages} {035} (\bibinfo {year} {2009})},\ \Eprint
  {http://arxiv.org/abs/0910.4481} {arXiv:0910.4481 [hep-ph]} \BibitemShut
  {NoStop}%
\bibitem [{\citenamefont {Gehrmann}\ \emph {et~al.}(2012)\citenamefont
  {Gehrmann}, \citenamefont {Jaquier}, \citenamefont {Glover},\ and\
  \citenamefont {Koukoutsakis}}]{Gehrmann:2011aa}%
  \BibitemOpen
  \bibfield  {author} {\bibinfo {author} {\bibfnamefont {T.}~\bibnamefont
  {Gehrmann}}, \bibinfo {author} {\bibfnamefont {M.}~\bibnamefont {Jaquier}},
  \bibinfo {author} {\bibfnamefont {E.~W.~N.}\ \bibnamefont {Glover}}, \ and\
  \bibinfo {author} {\bibfnamefont {A.}~\bibnamefont {Koukoutsakis}},\ }\href
  {\doibase 10.1007/JHEP02(2012)056} {\bibfield  {journal} {\bibinfo  {journal}
  {JHEP}\ }\textbf {\bibinfo {volume} {02}},\ \bibinfo {pages} {056} (\bibinfo
  {year} {2012})},\ \Eprint {http://arxiv.org/abs/1112.3554} {arXiv:1112.3554
  [hep-ph]} \BibitemShut {NoStop}%
\bibitem [{\citenamefont {Chen}\ \emph {et~al.}(2015)\citenamefont {Chen},
  \citenamefont {Gehrmann}, \citenamefont {Glover},\ and\ \citenamefont
  {Jaquier}}]{Chen:2014gva}%
  \BibitemOpen
  \bibfield  {author} {\bibinfo {author} {\bibfnamefont {X.}~\bibnamefont
  {Chen}}, \bibinfo {author} {\bibfnamefont {T.}~\bibnamefont {Gehrmann}},
  \bibinfo {author} {\bibfnamefont {E.~W.~N.}\ \bibnamefont {Glover}}, \ and\
  \bibinfo {author} {\bibfnamefont {M.}~\bibnamefont {Jaquier}},\ }\href
  {\doibase 10.1016/j.physletb.2014.11.021} {\bibfield  {journal} {\bibinfo
  {journal} {Phys. Lett. B}\ }\textbf {\bibinfo {volume} {740}},\ \bibinfo
  {pages} {147} (\bibinfo {year} {2015})},\ \Eprint
  {http://arxiv.org/abs/1408.5325} {arXiv:1408.5325 [hep-ph]} \BibitemShut
  {NoStop}%
\bibitem [{\citenamefont {Chen}\ \emph {et~al.}(2016)\citenamefont {Chen},
  \citenamefont {Cruz-Martinez}, \citenamefont {Gehrmann}, \citenamefont
  {Glover},\ and\ \citenamefont {Jaquier}}]{Chen:2016zka}%
  \BibitemOpen
  \bibfield  {author} {\bibinfo {author} {\bibfnamefont {X.}~\bibnamefont
  {Chen}}, \bibinfo {author} {\bibfnamefont {J.}~\bibnamefont {Cruz-Martinez}},
  \bibinfo {author} {\bibfnamefont {T.}~\bibnamefont {Gehrmann}}, \bibinfo
  {author} {\bibfnamefont {E.~W.~N.}\ \bibnamefont {Glover}}, \ and\ \bibinfo
  {author} {\bibfnamefont {M.}~\bibnamefont {Jaquier}},\ }\href {\doibase
  10.1007/JHEP10(2016)066} {\bibfield  {journal} {\bibinfo  {journal} {JHEP}\
  }\textbf {\bibinfo {volume} {10}},\ \bibinfo {pages} {066} (\bibinfo {year}
  {2016})},\ \Eprint {http://arxiv.org/abs/1607.08817} {arXiv:1607.08817
  [hep-ph]} \BibitemShut {NoStop}%
\bibitem [{\citenamefont {Ahmed}\ \emph {et~al.}(2014)\citenamefont {Ahmed},
  \citenamefont {Mahakhud}, \citenamefont {Mathews}, \citenamefont {Rana},\
  and\ \citenamefont {Ravindran}}]{Ahmed:2014pka}%
  \BibitemOpen
  \bibfield  {author} {\bibinfo {author} {\bibfnamefont {T.}~\bibnamefont
  {Ahmed}}, \bibinfo {author} {\bibfnamefont {M.}~\bibnamefont {Mahakhud}},
  \bibinfo {author} {\bibfnamefont {P.}~\bibnamefont {Mathews}}, \bibinfo
  {author} {\bibfnamefont {N.}~\bibnamefont {Rana}}, \ and\ \bibinfo {author}
  {\bibfnamefont {V.}~\bibnamefont {Ravindran}},\ }\href {\doibase
  10.1007/JHEP08(2014)075} {\bibfield  {journal} {\bibinfo  {journal} {JHEP}\
  }\textbf {\bibinfo {volume} {08}},\ \bibinfo {pages} {075} (\bibinfo {year}
  {2014})},\ \Eprint {http://arxiv.org/abs/1405.2324} {arXiv:1405.2324
  [hep-ph]} \BibitemShut {NoStop}%
\bibitem [{\citenamefont {Gehrmann-De~Ridder}\ \emph
  {et~al.}(2005)\citenamefont {Gehrmann-De~Ridder}, \citenamefont {Gehrmann},\
  and\ \citenamefont {Glover}}]{Gehrmann-DeRidder:2005btv}%
  \BibitemOpen
  \bibfield  {author} {\bibinfo {author} {\bibfnamefont {A.}~\bibnamefont
  {Gehrmann-De~Ridder}}, \bibinfo {author} {\bibfnamefont {T.}~\bibnamefont
  {Gehrmann}}, \ and\ \bibinfo {author} {\bibfnamefont {E.~W.~N.}\ \bibnamefont
  {Glover}},\ }\href {\doibase 10.1088/1126-6708/2005/09/056} {\bibfield
  {journal} {\bibinfo  {journal} {JHEP}\ }\textbf {\bibinfo {volume} {09}},\
  \bibinfo {pages} {056} (\bibinfo {year} {2005})},\ \Eprint
  {http://arxiv.org/abs/hep-ph/0505111} {arXiv:hep-ph/0505111} \BibitemShut
  {NoStop}%
\bibitem [{\citenamefont {Currie}\ \emph {et~al.}(2013)\citenamefont {Currie},
  \citenamefont {Glover},\ and\ \citenamefont {Wells}}]{Currie:2013vh}%
  \BibitemOpen
  \bibfield  {author} {\bibinfo {author} {\bibfnamefont {J.}~\bibnamefont
  {Currie}}, \bibinfo {author} {\bibfnamefont {E.~W.~N.}\ \bibnamefont
  {Glover}}, \ and\ \bibinfo {author} {\bibfnamefont {S.}~\bibnamefont
  {Wells}},\ }\href {\doibase 10.1007/JHEP04(2013)066} {\bibfield  {journal}
  {\bibinfo  {journal} {JHEP}\ }\textbf {\bibinfo {volume} {04}},\ \bibinfo
  {pages} {066} (\bibinfo {year} {2013})},\ \Eprint
  {http://arxiv.org/abs/1301.4693} {arXiv:1301.4693 [hep-ph]} \BibitemShut
  {NoStop}%
\bibitem [{\citenamefont {Fox}\ \emph {et~al.}(2024)\citenamefont {Fox},
  \citenamefont {Glover},\ and\ \citenamefont {Marcoli}}]{Fox:2024bfp}%
  \BibitemOpen
  \bibfield  {author} {\bibinfo {author} {\bibfnamefont {E.}~\bibnamefont
  {Fox}}, \bibinfo {author} {\bibfnamefont {N.}~\bibnamefont {Glover}}, \ and\
  \bibinfo {author} {\bibfnamefont {M.}~\bibnamefont {Marcoli}},\ }\href
  {\doibase 10.1007/JHEP12(2024)225} {\bibfield  {journal} {\bibinfo  {journal}
  {JHEP}\ }\textbf {\bibinfo {volume} {12}},\ \bibinfo {pages} {225} (\bibinfo
  {year} {2024})},\ \Eprint {http://arxiv.org/abs/2410.12904} {arXiv:2410.12904
  [hep-ph]} \BibitemShut {NoStop}%
\bibitem [{\citenamefont {Braun-White}\ \emph
  {et~al.}(2023{\natexlab{a}})\citenamefont {Braun-White}, \citenamefont
  {Glover},\ and\ \citenamefont {Preuss}}]{Braun-White:2023sgd}%
  \BibitemOpen
  \bibfield  {author} {\bibinfo {author} {\bibfnamefont {O.}~\bibnamefont
  {Braun-White}}, \bibinfo {author} {\bibfnamefont {N.}~\bibnamefont {Glover}},
  \ and\ \bibinfo {author} {\bibfnamefont {C.~T.}\ \bibnamefont {Preuss}},\
  }\href {\doibase 10.1007/JHEP06(2023)065} {\bibfield  {journal} {\bibinfo
  {journal} {JHEP}\ }\textbf {\bibinfo {volume} {06}},\ \bibinfo {pages} {065}
  (\bibinfo {year} {2023}{\natexlab{a}})},\ \Eprint
  {http://arxiv.org/abs/2302.12787} {arXiv:2302.12787 [hep-ph]} \BibitemShut
  {NoStop}%
\bibitem [{\citenamefont {Braun-White}\ \emph
  {et~al.}(2023{\natexlab{b}})\citenamefont {Braun-White}, \citenamefont
  {Glover},\ and\ \citenamefont {Preuss}}]{Braun-White:2023zwd}%
  \BibitemOpen
  \bibfield  {author} {\bibinfo {author} {\bibfnamefont {O.}~\bibnamefont
  {Braun-White}}, \bibinfo {author} {\bibfnamefont {N.}~\bibnamefont {Glover}},
  \ and\ \bibinfo {author} {\bibfnamefont {C.~T.}\ \bibnamefont {Preuss}},\
  }\href {\doibase 10.1007/JHEP11(2023)179} {\bibfield  {journal} {\bibinfo
  {journal} {JHEP}\ }\textbf {\bibinfo {volume} {11}},\ \bibinfo {pages} {179}
  (\bibinfo {year} {2023}{\natexlab{b}})},\ \Eprint
  {http://arxiv.org/abs/2307.14999} {arXiv:2307.14999 [hep-ph]} \BibitemShut
  {NoStop}%
\bibitem [{\citenamefont {Gehrmann-De~Ridder}\ \emph
  {et~al.}(2014)\citenamefont {Gehrmann-De~Ridder}, \citenamefont {Gehrmann},
  \citenamefont {Glover},\ and\ \citenamefont
  {Heinrich}}]{Gehrmann-DeRidder:2014hxk}%
  \BibitemOpen
  \bibfield  {author} {\bibinfo {author} {\bibfnamefont {A.}~\bibnamefont
  {Gehrmann-De~Ridder}}, \bibinfo {author} {\bibfnamefont {T.}~\bibnamefont
  {Gehrmann}}, \bibinfo {author} {\bibfnamefont {E.~W.~N.}\ \bibnamefont
  {Glover}}, \ and\ \bibinfo {author} {\bibfnamefont {G.}~\bibnamefont
  {Heinrich}},\ }\href {\doibase 10.1016/j.cpc.2014.07.024} {\bibfield
  {journal} {\bibinfo  {journal} {Comput. Phys. Commun.}\ }\textbf {\bibinfo
  {volume} {185}},\ \bibinfo {pages} {3331} (\bibinfo {year} {2014})},\ \Eprint
  {http://arxiv.org/abs/1402.4140} {arXiv:1402.4140 [hep-ph]} \BibitemShut
  {NoStop}%
\bibitem [{\citenamefont {Buckley}\ \emph {et~al.}(2015)\citenamefont
  {Buckley}, \citenamefont {Ferrando}, \citenamefont {Lloyd}, \citenamefont
  {Nordstr\"om}, \citenamefont {Page}, \citenamefont {R\"ufenacht},
  \citenamefont {Sch\"onherr},\ and\ \citenamefont {Watt}}]{Buckley:2014ana}%
  \BibitemOpen
  \bibfield  {author} {\bibinfo {author} {\bibfnamefont {A.}~\bibnamefont
  {Buckley}}, \bibinfo {author} {\bibfnamefont {J.}~\bibnamefont {Ferrando}},
  \bibinfo {author} {\bibfnamefont {S.}~\bibnamefont {Lloyd}}, \bibinfo
  {author} {\bibfnamefont {K.}~\bibnamefont {Nordstr\"om}}, \bibinfo {author}
  {\bibfnamefont {B.}~\bibnamefont {Page}}, \bibinfo {author} {\bibfnamefont
  {M.}~\bibnamefont {R\"ufenacht}}, \bibinfo {author} {\bibfnamefont
  {M.}~\bibnamefont {Sch\"onherr}}, \ and\ \bibinfo {author} {\bibfnamefont
  {G.}~\bibnamefont {Watt}},\ }\href {\doibase 10.1140/epjc/s10052-015-3318-8}
  {\bibfield  {journal} {\bibinfo  {journal} {Eur. Phys. J. C}\ }\textbf
  {\bibinfo {volume} {75}},\ \bibinfo {pages} {132} (\bibinfo {year} {2015})},\
  \Eprint {http://arxiv.org/abs/1412.7420} {arXiv:1412.7420 [hep-ph]}
  \BibitemShut {NoStop}%
\bibitem [{\citenamefont {Spira}(1998)}]{Spira:1997dg}%
  \BibitemOpen
  \bibfield  {author} {\bibinfo {author} {\bibfnamefont {M.}~\bibnamefont
  {Spira}},\ }\href {\doibase
  10.1002/(SICI)1521-3978(199804)46:3<203::AID-PROP203>3.0.CO;2-4} {\bibfield
  {journal} {\bibinfo  {journal} {Fortsch. Phys.}\ }\textbf {\bibinfo {volume}
  {46}},\ \bibinfo {pages} {203} (\bibinfo {year} {1998})},\ \Eprint
  {http://arxiv.org/abs/hep-ph/9705337} {arXiv:hep-ph/9705337} \BibitemShut
  {NoStop}%
\bibitem [{\citenamefont {Vermaseren}\ \emph {et~al.}(1997)\citenamefont
  {Vermaseren}, \citenamefont {Larin},\ and\ \citenamefont {van
  Ritbergen}}]{Vermaseren_1997}%
  \BibitemOpen
  \bibfield  {author} {\bibinfo {author} {\bibfnamefont {J.}~\bibnamefont
  {Vermaseren}}, \bibinfo {author} {\bibfnamefont {S.}~\bibnamefont {Larin}}, \
  and\ \bibinfo {author} {\bibfnamefont {T.}~\bibnamefont {van Ritbergen}},\
  }\href {\doibase 10.1016/s0370-2693(97)00660-6} {\bibfield  {journal}
  {\bibinfo  {journal} {Physics Letters B}\ }\textbf {\bibinfo {volume}
  {405}},\ \bibinfo {pages} {327–333} (\bibinfo {year} {1997})}\BibitemShut
  {NoStop}%
\bibitem [{\citenamefont {Baglio}\ and\ \citenamefont
  {Djouadi}(2011)}]{Baglio:2010ae}%
  \BibitemOpen
  \bibfield  {author} {\bibinfo {author} {\bibfnamefont {J.}~\bibnamefont
  {Baglio}}\ and\ \bibinfo {author} {\bibfnamefont {A.}~\bibnamefont
  {Djouadi}},\ }\href {\doibase 10.1007/JHEP03(2011)055} {\bibfield  {journal}
  {\bibinfo  {journal} {JHEP}\ }\textbf {\bibinfo {volume} {03}},\ \bibinfo
  {pages} {055} (\bibinfo {year} {2011})},\ \Eprint
  {http://arxiv.org/abs/1012.0530} {arXiv:1012.0530 [hep-ph]} \BibitemShut
  {NoStop}%
\end{thebibliography}%

\end{document}